\documentstyle[aps, preprint]{revtex}
\def\beq{\begin{eqnarray}}
\def\eed{\end{eqnarray}}
\begin{document}
\draft
\title{Optical and transport properties in doped two-leg ladder
antiferromagnet}
\author{Jihong Qin and Yun Song}
\address{Department of Physics, Beijing Normal University,
Beijing 100875, China}
\author{Shiping Feng}
\address{Department of Physics, Beijing Normal University, Beijing
100875, China\\
The Key Laboratory of Beam Technology and Material Modification of
Ministry of Education, Beijing Normal University, Beijing 100875,
China\\
National Laboratory of Superconductivity, Academia Sinica, Beijing
100080, China}
\author{Wei Yeu Chen}
\address{Department of Physics, Tamkang University, Tamsui 25137,
Taiwan}
\maketitle
\begin{abstract}
Within the $t$-$J$ model, the optical and transport properties of
the doped two-leg ladder antiferromagnet are studied based on the
fermion-spin theory. It is shown that the optical and transport
properties of the doped two-leg ladder antiferromagnet are mainly
governed by the holon scattering. The low energy peak in the
optical conductivity is located at a finite energy, while the
resistivity exhibits a crossover from the high temperature
metallic-like behavior to the low temperature insulating-like
behavior, which are consistent with the experiments.
\end{abstract}


The undoped cuprate superconductors are typical Mott insulators
with the antiferromagnetic (AF) long-range-order (AFLRO) \cite{n1}.
A small amount of carrier doping to this Mott insulating state
drives the metal-insulator transition and directly results in the
superconducting transition at low temperatures for low carrier
dopings. In the underdoped and optimally doped regimes, the normal
state above superconducting transition temperature shows many
unusual properties in the sense that they do not fit in the
conventional Fermi-liquid theory, and these unusual normal state
properties of the cuprate superconductors are closely related to
the special microscopic conditions, {\it i.e.}, Cu ions situated in
a {\it square-planar} arrangement and bridged by oxygen ions, weak
coupling between neighboring layers, and doping in such a way that
the Fermi level lies near the middle of the Cu-O $\sigma^{*}$ bond
\cite{n1,n2}. One common feature of these cuprate compounds is the
{\it square-planar} Cu arrangement \cite{n1,n2}. It is believed
that the two-dimensional (2D) anisotropy is prominent in the
cuprate superconductors due to the layered perovskite structure,
and strong quantum fluctuations with suppression of AFLRO are key
aspects \cite{n1,n2}. However, it has been reported recently that
some copper oxide materials, such as Sr$_{14}$Cu$_{24}$O$_{41}$,
do not contain CuO$_{2}$ planes common to the cuprate
superconductors but consist of two-leg Cu$_{2}$O$_{3}$ ladders and
edge-sharing CuO$_{2}$ chains \cite{n21,n3}. Moreover, the
isovalent substitution of Ca for Sr increases the hole density on
the ladders by a transfer of preexisting holes in the charge
reservoir layers composed of CuO$_{2}$ chains, and then the two-leg
ladder copper oxide material Sr$_{14-x}$Ca$_{x}$Cu$_{24}$O$_{41}$
is superconductor under pressure \cite{n4}. Moreover, the
experimental data show that the normal state is far from the
standard Fermi-liquid behavior \cite{n3,n5}. The neutron scattering
and muon spin resonance measurements on the compound
Sr$_{14}$Cu$_{24}$O$_{41}$ indicate the system is the
antiferromagnet with the short-range spin order \cite{n3,n6},
while the transport measurements on the material
Sr$_{14-x}$Ca$_{x}$Cu$_{24}$O$_{41}$ show that the behavior of the
temperature dependent resistivity is characterized by a crossover
from the high temperature metallic-like to the low temperature
insulating-like \cite{n4,n5}, which may be in common with these
of the heavily underdoped cuprate superconducting materials
\cite{n1,n2,n61}. Further, it has been shown \cite{n4,n5} from the
experiments that the ratio of the interladder to in-ladder
resistivities is $R=\rho_{a}(T)/\rho_{c}(T)\sim 10$, this large
magnitude of the resistivity anisotropy reflects that the
interladder mean free path is shorter than the interladder
distance, and the carriers are tightly confined to the ladders, and
also is the evidence of the incoherent transport in the interladder
direction, therefore the common two-leg ladders in the ladder
materials clearly dominate the most normal state properties. These
ladder copper oxide materials also are natural extensions of the
Cu-O chain compounds towards the CuO$_{2}$ sheet structures. Other
two-leg ladder compounds have also been found \cite{n7,n3}, such as
experiments suggest the realization of the two-leg ladder
spin-one-half antiferromagnet in (VO)$_{2}$P$_{2}$O$_{7}$. On the
theoretical hand, the two-leg ladder antiferromagnet may therefore
be regarded as realizations of the unique, coherent resonating
valence bond spin liquid, which may play a crucial role in the
superconductivity of the cuprate superconductors as emphasized by
Anderson \cite{n8}. Therefore it is very important to investigate
the normal state properties of the doped two-leg ladder system by
a systematic approach since it may get deeper insights into the
still not fully understood anomalous normal state of the cuprate
superconductors.

Many researchers have argued successfully that the $t$-$J$ model,
acting on the Hilbert space with no doubly occupied site, provides
a consistent description of the physical properties of the doped
antiferromagnet \cite{n8,n9}. Within the $t$-$J$ model, the normal
state properties of the cuprate superconductors have been studied
\cite{n2,n10,n12}, and the results show that the unusual normal
state properties in the cuprate superconductors are caused by the
strong electron correlation. Since the strong electron correlation
is common for both cuprate superconductors and doped two-leg ladder
antiferromagnet, then the unconventional normal state properties in
the doped two-leg ladder antiferromagnet may be similar to these in
the cuprate superconductors. In this paper, we study the optical
and transport properties of the doped two-leg ladder
antiferromagnet within the $t$-$J$ model. Our results show that the
low energy peak in the optical conductivity is located at a finite
energy ($\omega\sim 0.2t$), while the resistivity exhibits a
crossover from the high temperature metallic-like behavior to the
low temperature insulating-like behavior.

The basic element of the two-leg ladder materials is the two-leg
ladder, which is defined as two parallel chains of ions, with bonds
among them such that the interchain coupling is comparable in
strength to the couplings along the chains, while the coupling
between the two chains that participates in this structure is
through rungs \cite{n3}. In this case, the $t$-$J$ model on the
two-leg ladder is expressed as,
\begin{eqnarray}
H&=&-t_{\parallel}\sum_{i\hat{\eta}a\sigma}C_{ia\sigma}^{\dagger}
C_{i+\hat{\eta}a\sigma}-t_{\perp}\sum_{i\sigma}
(C_{i1\sigma}^{\dagger}C_{i2\sigma}+{\rm h.c.}) \nonumber \\
&-&\mu\sum_{ia\sigma}C_{ia\sigma }^{\dagger}C_{ia\sigma }+
J_{\parallel}\sum_{i\hat{\eta}a}{\bf S}_{ia}\cdot
{\bf S}_{i+\hat{\eta}a}+J_{\perp}\sum_{i}{\bf S}_{i1}\cdot
{\bf S}_{i2},
\end{eqnarray}
where $\hat{\eta}=\pm c_{0}$, $c_{0}$ is the lattice constant of
the two-leg ladder lattice, which is set as the unit hereafter,
$i$ runs over all rungs, $\sigma(=\uparrow,\downarrow)$ and
$a(=1,2)$ are spin and leg indices, respectively,
$C^{\dagger}_{ia\sigma}$ ($C_{ia\sigma}$) are the electron
creation (annihilation) operators, ${\bf S}_{ia}=C^{\dagger}_{ia}
\vec{\sigma}C_{ia}/2$ are the spin operators with $\vec{\sigma}=
(\sigma_{x},\sigma_{y},\sigma_{z})$ as the Pauli matrices, and
$\mu$ is the chemical potential. This $t$-$J$ model is supplemented
by the on-site single occupancy local constraint $\sum_{\sigma}
C_{ia\sigma}^{\dagger}C_{ia\sigma}\leq 1$. The two-leg ladder with
$c$ and $a$ axes parallel to ladders and rungs, respectively, is
sketched in Fig. 1. In the materials of interest, the exchange
coupling $J_{\parallel}$ along the legs is nearly the same as the
exchange coupling $J_{\perp}$ across a rung, and similarly the
hopping $t_{\parallel}$ along the legs is close to the rung hopping
strength $t_{\perp}$, therefore, in the following discussions, we
will work with the isotropic system, $J_{\perp}=J_{\parallel}=J$,
$t_{\perp}=t_{\parallel}=t$.

In the $t$-$J$ model, the strong electron correlation is reflected
by the local constraint. To incorporate this local constraint, the
fermion-spin theory based on the charge-spin separation,
$C_{ia\uparrow}=h_{ia}^{\dagger}S_{ia}^{-}$, $C_{ia\downarrow}=
h_{ia}^{\dagger}S_{ia}^{+}$, has been proposed \cite{n11}, where
the spinless fermion operator $h_{ia}$ keeps track of the charge
(holon), while the pseudospin operator $S_{ia}$ keeps track of the
spin (spinon), and then the local constraint can be treated
properly in analytical calculations. Within this fermion-spin
representation, the low-energy behavior of the $t$-$J$ model (1)
can be rewritten as,
\begin{eqnarray}
H&=&t\sum_{i\hat{\eta}a}h_{i+\hat{\eta}a}^{\dagger}h_{ia}
(S_{ia}^{+}S_{i+\hat{\eta}a}^{-}+S_{ia}^{-}S_{i+\hat{\eta}a}^{+})
+t\sum_{i}(h_{i1}^{\dagger}h_{i2}+h_{i2}^{\dagger}h_{i1})
(S_{i1}^{+}S_{i2}^{-}+S_{i1}^{-}S_{i2}^{+}) \nonumber \\
&+&\mu\sum_{ia}h_{ia}^{\dagger}h_{ia}+J_{\parallel {\rm eff}}
\sum_{i\hat{\eta}a}{\bf S}_{ia}\cdot {\bf S}_{i+\hat{\eta}a}+
J_{\perp {\rm eff}}\sum_{i}{\bf S}_{i1}\cdot {\bf S}_{i2},
\end{eqnarray}
where
$J_{\parallel {\rm eff}}=J[(1-\delta)^{2}-\phi^{2}_{\parallel}]$,
$J_{\perp {\rm eff}}=J[(1-\delta)^{2}-\phi^{2}_{\perp}]$, the holon
particle-hole order parameters $\phi_{\parallel}=\langle
h^{\dagger}_{ia}h_{i+\hat{\eta}a}\rangle$, $\phi_{\perp}=\langle
h^{\dagger}_{i1}h_{i2}\rangle$, $\delta$ is the hole doping
concentration, and $S^{+}_{i}$ and $S^{-}_{i}$ are the pseudospin
raising and lowering operators, respectively. Since the local
constraint has been treated properly in the framework of the
fermion-spin theory, then the extra gauge degree of freedom related
with the electron on-site local constraint under the charge-spin
separation does not appear. In this case, the spin fluctuation
couples only to spinons, while the charge fluctuation couples only
to holons, but the strong correlation between holons and spinons
is still considered through the holon's order parameters entering
the spinon's propagator and the spinon's order parameters entering
the holon's propagator, therefore both holons and spinons
contribute to the charge dynamics. In this case, the optical and
transport properties of the doped cupares have been discussed
\cite{n12}, and the results are consistent with the experiments
\cite{n1,n2}. Following their discussions \cite{n12}, the optical
conductivity of the doped two-leg ladder antiferromagnet can be
expressed as,
\begin{eqnarray}
\sigma_{c}(\omega)=-{{\rm Im}\Pi^{(h)}(\omega)\over \omega},
\end{eqnarray}
with $\Pi^{(h)}(\omega)$ is the holon current-current correlation
function, which are defined as, $\Pi^{(h)}(\tau-\tau')=-\langle
T_\tau j^{(h)}(\tau)j^{(h)}(\tau')\rangle$, where $\tau$ and
$\tau'$ are the imaginary times, and $T_{\tau}$ is the $\tau$ order
operator. Within the $t$-$J$ Hamiltonian (2), the current densities
of holons is obtained by the time derivation of the polarization
operator using Heisenberg's equation of motion as,
$j^{(h)}=2\chi_{\parallel}et\sum_{ai\hat{\eta}}\hat{\eta}
h_{ai+\hat{\eta}}^{\dagger} h_{ai}+2\chi_{\perp}et\sum_{i}(R_{2i}-
R_{1i})(h^{\dagger}_{2i}h_{1i}-h^{\dagger}_{1i}h_{2i})$, where
$R_{1i}$ and $R_{2i}$ are lattice sites of the leg $1$ and leg
$2$, respectively, the spinon correlation functions
$\chi_{\parallel}=\langle S_{ai}^{+}S_{ai+\hat{\eta}}^{-}\rangle$,
$\chi_{\perp}=\langle S^{+}_{1i}S^{-}_{2i}\rangle$, and $e$ is the
electronic charge, which is set as the unit hereafter. This holon
current-current correlation function can be calculated in terms of
the holon Green's function $g(k,\omega)$. However, in the two-leg
ladder system, because there are two coupled $t$-$J$ chains, then
the energy spectrum has two branches. In this case, the
one-particle holon Green's function is the matrix, and can be
expressed as $g(i-j,\tau-\tau')=g_{L}(i-j,\tau-\tau')+ \sigma_{x}
g_{T}(i-j,\tau-\tau')$, where the longitudinal and transverse parts
are defined as $g_{L}(i-j,\tau-\tau')=-\langle T_{\tau}h_{ai}(\tau)
h^{\dagger}_{aj}(\tau')\rangle$ and $g_{T}(i-j,\tau-\tau')=-\langle
T_{\tau}h_{ai}(\tau)h^{\dagger}_{a'j}(\tau')\rangle$ ($a\neq a'$),
respectively. Then after a straightforward calculation \cite{n12},
we obtain the optical conductivity of the doped two-leg ladder
antiferromagnet as,
\begin{eqnarray}
\sigma_{c}(\omega)=\sigma^{(L)}_{c}(\omega)+\sigma^{(T)}_{c}
(\omega),
\end{eqnarray}
with the longitudinal and transverse parts are given by,
\begin{mathletters}
\begin{eqnarray}
\sigma^{(L)}_{c}(\omega)&=&4t^{2}{1\over L}\sum_{k}
(4\chi^{2}_{\parallel}{\rm sin}^{2}k+\chi^{2}_{\perp})
\int^{\infty}_{-\infty}{d\omega'\over 2\pi}A^{(h)}_{L}
(k,\omega'+\omega)A^{(h)}_{L}(k,\omega'){n_{F}(\omega'+\omega)-
n_{F}(\omega') \over \omega}, \\
\sigma^{(T)}_{c}(\omega)&=&4t^{2}{1\over L}\sum_{k}
(4\chi^{2}_{\parallel}{\rm sin}^{2}k-\chi^{2}_{\perp})
\int^{\infty}_{-\infty}{d\omega'\over 2\pi}A^{(h)}_{T}
(k,\omega'+\omega)A^{(h)}_{T}(k,\omega'){n_{F}(\omega'+\omega)-
n_{F}(\omega') \over \omega},
\end{eqnarray}
\end{mathletters}
respectively, where $L$ is the number of rungs, $n_{F}(\omega)$ is
the fermion distribution function, and the longitudinal and
transverse holon spectral function $A^{(h)}_{L}(k,\omega)$ and
$A^{(h)}_{T}(k,\omega)$ are obtained as, $A^{(h)}_{L}(k,\omega)
=-2{\rm Im}g_{L}(k,\omega)$ and $A^{(h)}_{T}(k,\omega)=-2{\rm Im}
g_{T}(k,\omega)$, respectively, the full holon Green's function
$g^{-1}(k,\omega)=g^{(0)-1}(k,\omega)-\Sigma^{(h)}(k,\omega)$ with
the longitudinal and transverse mean-field holon Green's function
$g^{(0)}_{L}(k,\omega)=1/2\sum_{\nu}1/(\omega-\xi^{(\nu)}_{k})$
and $g^{(0)}_{T}(k,\omega)=1/2\sum_{\nu}(-1)^{\nu+1}1/(\omega-
\xi^{(\nu)}_{k})$, where $\nu=1,2$, while the longitudinal and
transverse second-order holon self-energy from the spinon pair
bubble are obtained by the loop expansion to the second-order
\cite{n12} as,
\begin{eqnarray}
\Sigma_{L}(k,\omega)&=&({t\over N})^{2}\sum_{pq}\sum_{\nu\nu'\nu''}
\Xi_{\nu\nu'\nu''}(k,p,q,\omega), \\
\Sigma_{T}(k,\omega)&=&({t\over N})^{2}\sum_{pq}\sum_{\nu\nu'\nu''}
(-1)^{\nu+\nu'+\nu''+1}\Xi_{\nu\nu'\nu''}(k,p,q,\omega),
\end{eqnarray}
respectively, with $\Xi_{\nu\nu'\nu''}(k,p,q,\omega)$ is given by,
\begin{eqnarray}
\Xi_{\nu\nu'\nu''}(k,p,q,\omega)={B^{(\nu')}_{q+p}B^{(\nu)}_{q}
\over 32\omega^{(\nu')}_{q+p}\omega^{(\nu)}_{q}}\left
(2[\gamma_{q+p+k}+\gamma_{q-k}]+[(-1)^{\nu+\nu''}+(-1)^{\nu'+\nu''}]
\right )^{2}\left ({F^{(1)}_{\nu\nu'\nu''}(k,p,q)\over\omega+
\omega^{(\nu')}_{q+p}-\omega^{(\nu)}_{q}-\xi^{(\nu'')}_{p+k}}
\right. \nonumber \\
\left. +{F^{(2)}_{\nu\nu'\nu''}(k,p,q)\over\omega-
\omega^{(\nu')}_{q+p}+\omega^{(\nu)}_{q}-\xi^{(\nu'')}_{p+k}}+
{F^{(3)}_{\nu\nu'\nu''}(k,p,q)\over\omega+\omega^{(\nu')}_{q+p}+
\omega^{(\nu)}_{q}-\xi^{(\nu'')}_{p+k}}+{F^{(4)}_{\nu\nu'\nu''}
(k,p,q)\over\omega-\omega^{(\nu')}_{q+p}-\omega^{(\nu)}_{q}-
\xi^{(\nu'')}_{p+k}}\right ),~~~~~~
\end{eqnarray}
where $\gamma_{k}={\rm cos}k$, $\lambda=4J_{\parallel{\rm eff}}$,
$\epsilon_{\parallel}=1+2t\phi_{\parallel}/J_{\parallel{\rm eff}}$,
$\epsilon_{\perp}=1+4t\phi_{\perp}/J_{\perp{\rm eff}}$, and
\begin{mathletters}
\begin{eqnarray}
B^{(\nu)}_{k}&=&B_{k}-J_{\perp {\rm eff}}[\chi_{\perp}+
2\chi^{z}_{\perp}(-1)^{\nu}][\epsilon_{\perp}+(-1)^{\nu}],\\
B_{k}&=&\lambda [(2\epsilon_{\parallel}\chi^{z}_{\parallel}+
\chi_{\parallel})\gamma_{k}-(\epsilon_{\parallel}\chi_{\parallel}
+2\chi^{z}_{\parallel})], \\
F^{(1)}_{\nu\nu'\nu''}(k,p,q)&=&n_{F}(\xi^{(\nu'')}_{p+k})[n_{B}
(\omega^{(\nu)}_{q})-n_{B}(\omega^{(\nu')}_{q+p})]+n_{B}
(\omega^{(\nu')}_{q+p})[1+n_{B}(\omega^{(\nu)}_{q})], \\
F^{(2)}_{\nu\nu'\nu''}(k,p,q)&=&n_{F}(\xi^{(\nu'')}_{p+k})[n_{B}
(\omega^{(\nu')}_{q+p})-n_{B}(\omega^{(\nu)}_{q})]+n_{B}
(\omega^{(\nu)}_{q})[1+n_{B}(\omega^{(\nu')}_{q+p})],\\
F^{(3)}_{\nu\nu'\nu''}(k,p,q)&=&n_{F}(\xi^{(\nu'')}_{p+k})[1+n_{B}
(\omega^{(\nu')}_{q+p})+n_{B}(\omega^{(\nu)}_{q})]+n_{B}
(\omega^{(\nu)}_{q})n_{B}(\omega^{(\nu')}_{q+p}),\\
F^{(4)}_{\nu\nu'\nu''}(k,p,q)&=&[1+n_{B}(\omega^{(\nu)}_{q})]
[1+n_{B}(\omega^{(\nu')}_{q+p})]-n_{F}(\xi^{(\nu'')}_{p+k})[1+
n_{B}(\omega^{(\nu')}_{q+p})+n_{B}(\omega^{(\nu)}_{q})],
\end{eqnarray}
\end{mathletters}
with $n_{B}(\omega^{(\nu)}_{k})$ is the boson distribution
functions, the MF holon excitations $\xi^{(\nu)}_{k}=4t
\chi_{\parallel}\gamma_{k}+\mu+2\chi_{\perp}t(-1)^{\nu+1}$, and
the MF spinon excitations,
\begin{eqnarray}
\omega^{(\nu)2}_{k}&=&\alpha\epsilon_{\parallel}\lambda^{2}
({1\over 2}\chi_{\parallel}+\epsilon_{\parallel}
\chi^{z}_{\parallel})\gamma_{k}^{2}-\epsilon_{\parallel}\lambda^{2}
[{1\over 2}\alpha({1\over 2}\epsilon_{\parallel}\chi_{\parallel}+
\chi^{z}_{\parallel})+\alpha(C^{z}_{\parallel}+{1\over 2}
C_{\parallel})+{1\over 4}(1-\alpha)]\gamma_{k} \nonumber \\
&-&{1\over 2}\alpha\epsilon_{\perp}\lambda J_{\perp{\rm eff}}
(C_{\perp}+\epsilon_{\parallel}\chi_{\perp})\gamma_{k}+\alpha
\lambda J_{\perp{\rm eff}}(-1)^{\nu+1}[{1\over 2}(\epsilon_{\perp}
\chi_{\parallel}+\epsilon_{\parallel}\chi_{\perp})+
\epsilon_{\parallel}\epsilon_{\perp}(\chi^{z}_{\perp}+
\chi^{z}_{\parallel})]\gamma_{k} \nonumber \\
&-&\alpha\epsilon_{\parallel}\lambda J_{\perp{\rm eff}}
(C^{z}_{\perp}+\chi^{z}_{\perp})\gamma_{k}+\lambda^{2}[\alpha
(C^{z}_{\parallel}+{1\over 2}\epsilon^{2}_{\parallel}C_{\parallel})
+{1\over 8}(1-\alpha)(1+\epsilon^{2}_{\parallel})-{1\over 2}
\alpha\epsilon_{\parallel}({1\over 2}\chi_{\parallel}+
\epsilon_{\parallel}\chi^{z}_{\parallel})] \nonumber \\
&+&\alpha\lambda J_{\perp{\rm eff}}[\epsilon_{\parallel}
\epsilon_{\perp}C_{\perp}+2C^{z}_{\perp}]+{1\over 4}
J^{2}_{\perp{\rm eff}}(\epsilon^{2}_{\perp}+1)-{1\over 2}
\epsilon_{\perp}J^{2}_{\perp{\rm eff}}(-1)^{\nu+1} \nonumber \\
&-&\alpha\lambda J_{\perp{\rm eff}}(-1)^{\nu+1}[{1\over 2}
\epsilon_{\parallel}\epsilon_{\perp}\chi_{\parallel}+
\epsilon_{\perp}(\chi^{z}_{\parallel}+C^{z}_{\perp})+{1\over 2}
\epsilon_{\parallel}C_{\perp}],
\end{eqnarray}
with the spinon correlation functions $\chi^{z}_{\parallel}=
\langle S_{ai}^{z}S_{ai+\hat{\eta}}^{z}\rangle$, $\chi^{z}_{\perp}=
\langle S_{1i}^{z}S_{2i}^{z}\rangle$, $C_{\parallel}=(1/4)
\sum_{\hat{\eta}\hat{\eta'}}\langle S_{ai+\hat{\eta}}^{+}
S_{ai+\hat{\eta'}}^{-}\rangle$, and $C^{z}_{\parallel}=(1/4)
\sum_{\hat{\eta}\hat{\eta'}}\langle S_{ai+\hat{\eta}}^{z}
S_{ai+\hat{\eta'}}^{z}\rangle$, $C_{\perp}=(1/2)\sum_{\hat{\eta}}
\langle S_{2i}^{+}S_{1i+\hat{\eta}}^{-}\rangle$, and
$C^{z}_{\perp}=(1/2)\sum_{\hat{\eta}}\langle S_{1i}^{z}S_{2i+
\hat{\eta}}^{z}\rangle$. In order not to violate the sum rule of
the correlation function $\langle S^{+}_{ai}S^{-}_{ai}\rangle=1/2$
in the case without AFLRO, the important decoupling parameter
$\alpha$ has been introduced in the mean-field calculation, which
can be regarded as the vertex correction \cite{n13}. All the
above mean-field order parameters have been determined by the
self-consistent calculation \cite{n13}.

Now we discuss the optical and transport properties of the doped
two-leg ladder antiferromagnet. The optical conductivity of the
doped Mott insulator in principle consists of three different
pieces: (1) the Drude absorption, (2) absorption across the
Mott-Hubbard gap which rapidly decreases in intensity upon doping,
and (3) an absorption continuum within the gap, which reflects the
strong coupling between charge carriers and spin excitations. In
Fig. 2, we plot the optical conductivity $\sigma_{c}(\omega)$ at
doping (a) $\delta=0.16$ and (b) $\delta=0.20$ for parameter
$t/J=2.5$ with temperature $T=0$ in comparison with the
corresponding experimental data \cite{n5} taken on
Sr$_{14-x}$Ca$_{x}$Cu$_{24}$O$_{41}$ (inset), where the hole
density on the ladders in $x=8$ and $x=11$ is $\delta=0.16$ and
$\delta=0.20$ per ladder Cu, respectively. From Fig. 2, we find
the optical conductivity consists of two bands separated at
$\omega\sim 0.5t$, the higher-energy band, corresponding to the
midinfrared band, shows a weak peak at $\omega\sim 0.8t$, unlike
a Drude peak which dominates in the conductivity spectrum of the
doped cuprate superconductors, the lower-energy peak in the
present ladder systems is located at a finite energy
$\omega\sim 0.2t$, while the extremely small continuum absorption
is consistent with the notion of the charge-spin separation.
These behaviors are in agreement with the experimental results of
the doped two-leg ladder antiferromagnet \cite{n4,n5}. In the
above calculations, we also find that the optical conductivity
$\sigma_{c}(\omega)$ of the doped two-leg ladder antiferromagnet
is essentially determined by its longitudinal part
$\sigma^{(L)}_{c}(\omega)$, this is why in the present ladder
systems the midinfrared band is much weaker than the low energy
band, and conductivity spectrum appears to reflect the
one-dimensional nature of the electronic state \cite{n14}. This
conductivity of the doped two-leg ladder antiferromagnet has been
discussed by Kim \cite{n145} based on a model of hole pairs
forming a strongly correlated liquid, where quantum interference
effects are handled using renormalization group methods, and then
the main low-energy features of the experiments are reproduced.
Our results in low-energy are also consistent with his results.

With the help of the optical conductivity (4), the resistivity can
be obtained as $\rho_{c}=1/\lim_{\omega\rightarrow 0}\sigma_{c}
(\omega)$. The result of $\rho_{c}$ at doping $\delta=0.16$ (solid
line) and $\delta=0.20$ (dashed line) for parameter $t/J=2.5$ is
shown in Fig. 3 in comparison with the corresponding experimental
results \cite{n5} taken on Sr$_{14-x}$Ca$_{x}$Cu$_{24}$O$_{41}$
(inset). Our results show that the behavior of the temperature
dependence of $\rho_{c}(T)$ exhibits a crossover from the high
temperature metallic-like to the low temperature insulating-like,
but the metallic-like temperature dependence dominates over a wide
temperature range, in agreement with the corresponding experimental
data \cite{n4,n5}. The present result also indicates that the
behaviors of $\rho_{c}(T)$ in the doped two-leg ladder
antiferromagnet are very similar to these of the heavily
underdoped cuprate superconducting materials \cite{n1,n2,n61}
perhaps since both materials have almost same microscopic energy
scales and owe to the common corner-sharing CuO$_{4}$ networks.

In the above discussions, the central concern of the optical and
transport properties in the doped two-leg ladder antiferromagnet
is the quasi-one dimensionality of the electron state, then the
optical and transport properties are mainly determined by the
longitudinal charged holon fluctuation. Our present study also
indicates that the observed crossovers of $\rho_{c}$ for the doped
two-leg ladder antiferromagnet seems to be connected with the
pseudogap in the charge holon excitations, which can be understood
from the physical property of the holon density of states (DOS),
$\Omega_{L}(\omega)=1/N\sum_{k}A^{(h)}_{L}(k,\omega)$. This holon
DOS has been calculated, and the result at doping $\delta=0.16$
for parameter $t/J=2.5$ with temperature $T=0$ is plotted in
Fig. 4. We therefore find that the holon DOS consists of a U-shape
pseudogap near the chemical potential $\mu$. For the better
understanding of the property of this U-shape pseudogap, we plot
the phase diagram $T^{*}\sim \delta$ at parameter $t/J=2.5$ in
Fig. 5, where $T^{*}$ marks the development of the pseudogap in
the holon DOS. As seen from Fig. 5, this pseudogap is doping and
temperature dependent, and grows monotonously as the doping
$\delta$ decreases, and disappear in higher doping. Moreover, this
pseudogap decreases with increasing temperatures, and vanishes at
higher temperatures. Since the full holon Green's function (then
the holon spectral function and DOS) is obtained by considering
the second-order correction due to the spinon pair bubble, then
the holon pseudogap is closely related to the spinon fluctuation.
This holon pseudogap would reduce the holon scattering and thus
is responsible for the metallic to insulating crossover in the
resistivity $\rho_{c}$. While in the region where the holon
pseudogap closes at high temperatures, the charged holon
scattering would give rise to the metallic temperature dependence
of the resistivity.

In summary, we have studied the optical and transport properties
of the doped two-leg ladder antiferromagnet within the $t$-$J$
model. Our result shows the optical and transport properties of
the doped two-leg ladder antiferromagnet are mainly governed by
the charged holon scattering. The low energy peak in the optical
conductivity is located at a finite energy, while the resistivity
exhibits a crossover from the high temperature metallic-like
behavior to the low temperature insulating-like behavior, in
agreement with the experiments.

Finally, we emphasize that in the above discussions, only the
results of the doped isotropic two-leg ladder system are presented.
However, we \cite{n146} have also studied the physical properties
of the anisotropic system, i.e., $J_{\perp}<J_{\parallel}$ and
$t_{\perp}<t_{\parallel}$. In this case, the interference effects
between two legs are decreased with decreasing the values of
$J_{\perp}/J_{\parallel}$ and $t_{\perp}/t_{\parallel}$, this
leads to that the low-energy peak of the conductivity is located
at $\omega\sim 0$ instead of a finite energy for the isotropic
system. On the other hand, it has been shown that the interleg
single-electron hopping changes the asymptotic behavior of the
interleg spin-spin correlation functions, but their exponents are
independent of the interleg coupling strength \cite{n147}. We
believe that the evolution of the incommensurate magnetic
fluctuations with dopings in the doped square lattice
antiferromagnet \cite{n148} will also occur in the doped two-leg
antiferromagnet, and the related theoretical results will be
presented elsewhere.

\acknowledgments
The authors would like to thank Dr. Feng Yuan and Xianglin Ke for
helpful discussions. This work was supported by the National
Natural Science Foundation for Distinguished Young Scholars under
Grant No. 10125415, the National Natural Science Foundation under
Grant No. 10074007, 90103024, the Grant from Ministry of Education
of China, and the National Science Council under Grant No. NSC
90-2816-M-032-0001-6.

\begin{figure}
\caption{The $t$-$J$ ladder with two legs and $L$ rungs. The
couplings along the legs are $t_{\parallel}$ and $J_{\parallel}$,
and those along the rungs $t_{\perp}$ and $J_{\perp}$.}
\end{figure}

\begin{figure}
\caption{The optical conductivity of the doped two-leg ladder
antiferromagnet at doping (a) $\delta=0.16$ and (b) $\delta=0.20$
in parameter $t/J=2.5$ with temperature $T=0$. Inset: the
experimental result on Sr$_{14-x}$Ca$_{x}$Cu$_{24}$O$_{41}$ taken
from Ref. [6].}
\end{figure}

\begin{figure}
\caption{The resistivity of the doped two-leg ladder
antiferromagnet at doping $\delta=0.16$ (solid line) and
$\delta=0.20$ (dashed line) in parameter $t/J=2.5$. Inset: the
experimental result on Sr$_{14-x}$Ca$_{x}$Cu$_{24}$O$_{41}$ taken
from Ref. [6].}
\end{figure}

\begin{figure}
\caption{The holon density of states at doping $\delta=0.16$ in
parameter $t/J=2.5$ with temperature $T=0$.}
\end{figure}

\begin{figure}
\caption{The normal-state phase diagram $T^{*}\sim \delta$ for
parameter $t/J=2.5$. $T^{*}$ marks the development of the holon
pseudogap in the holon density of states}
\end{figure}

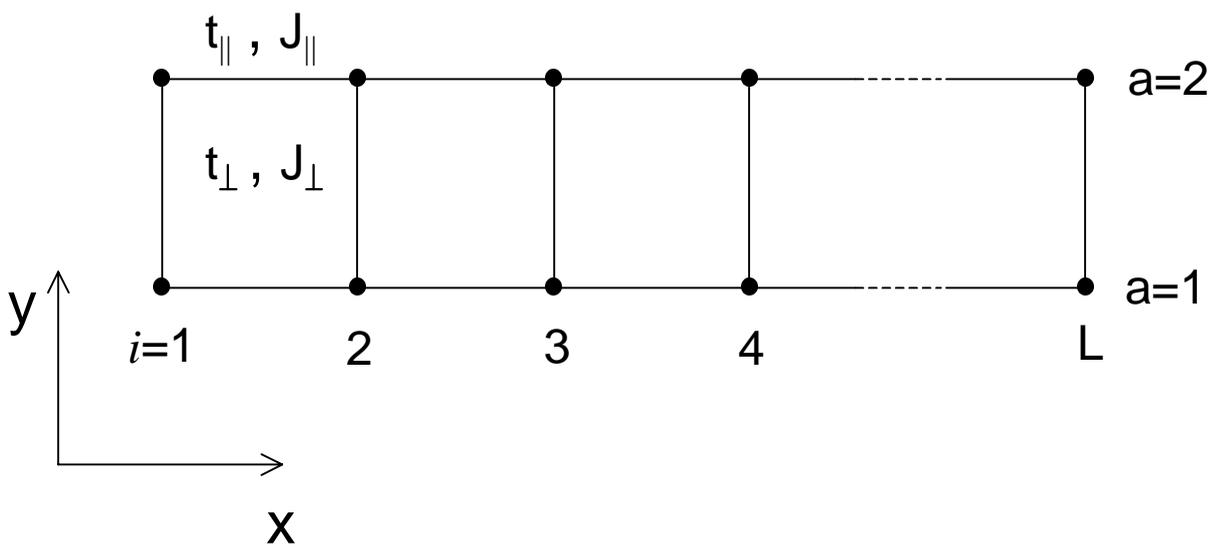
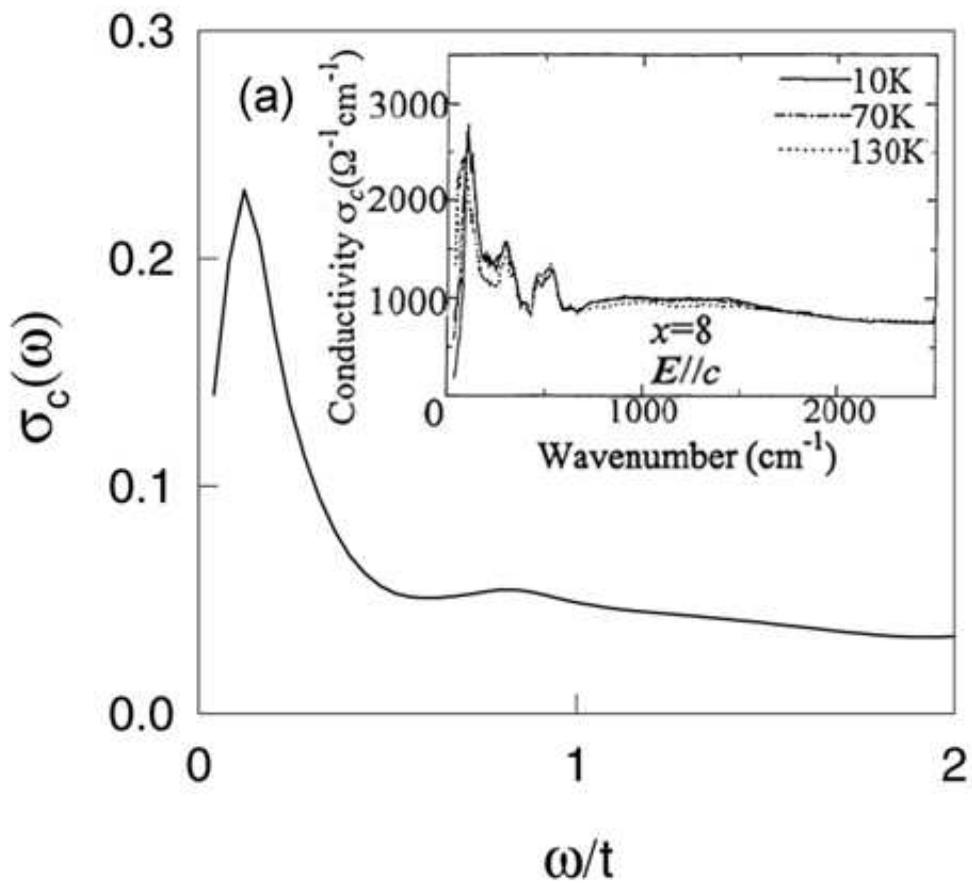
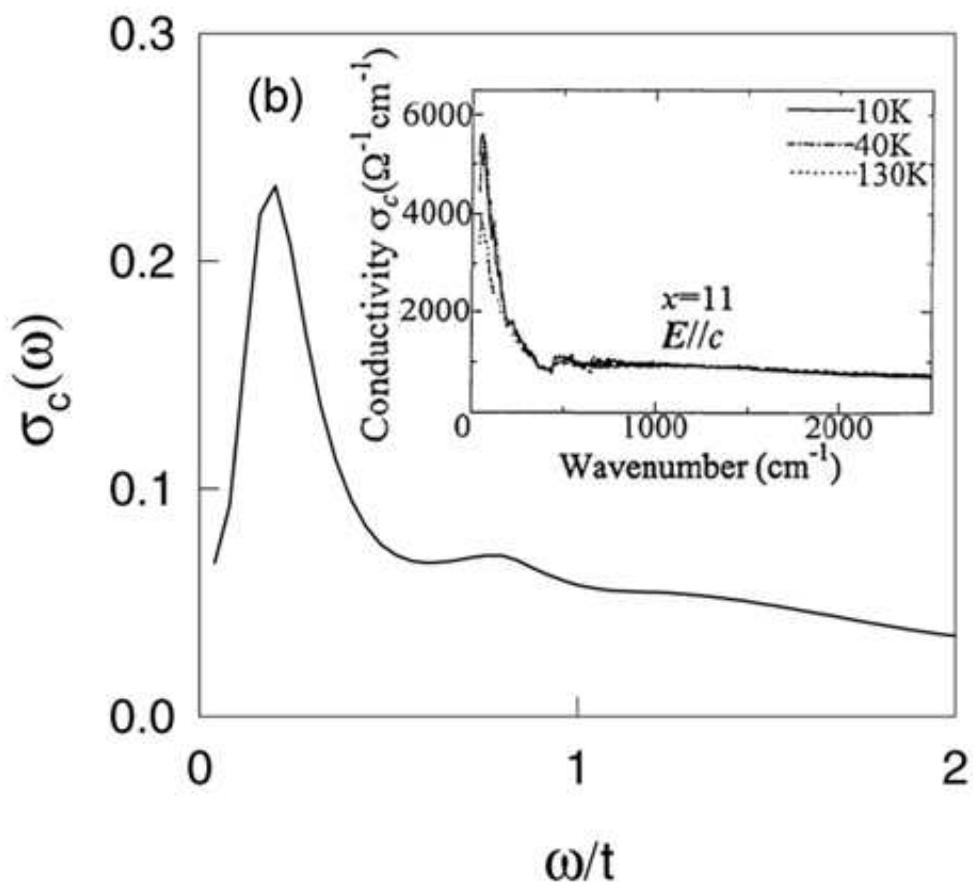
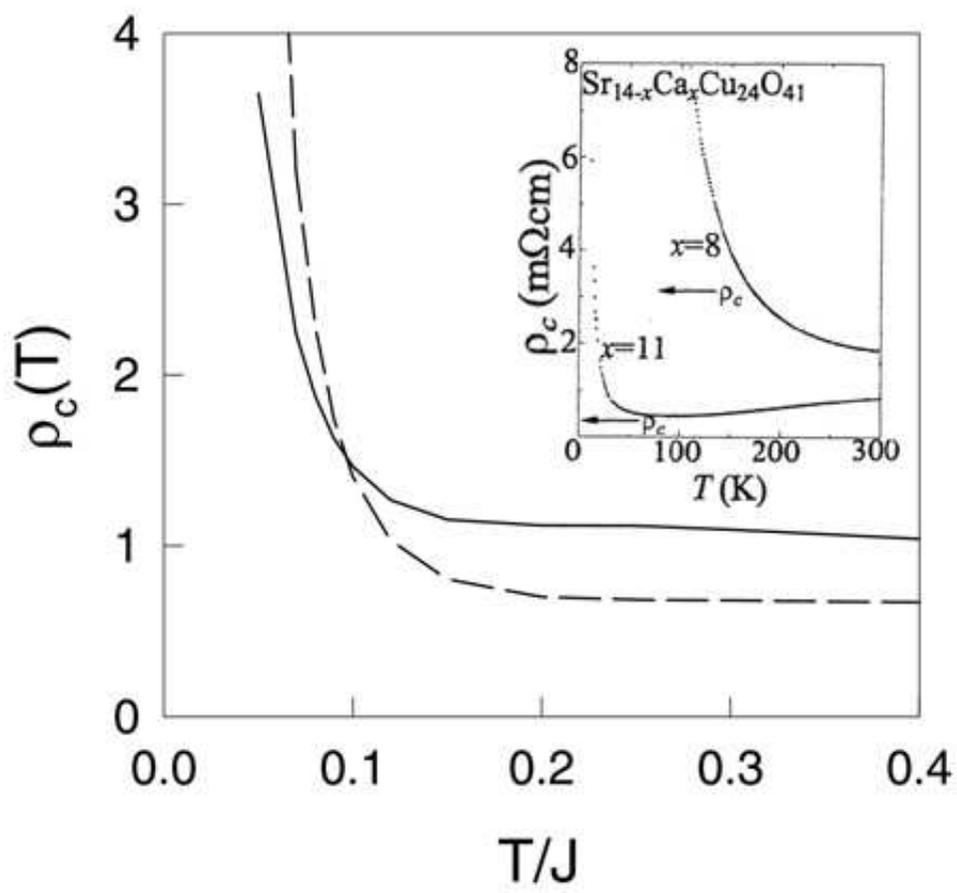
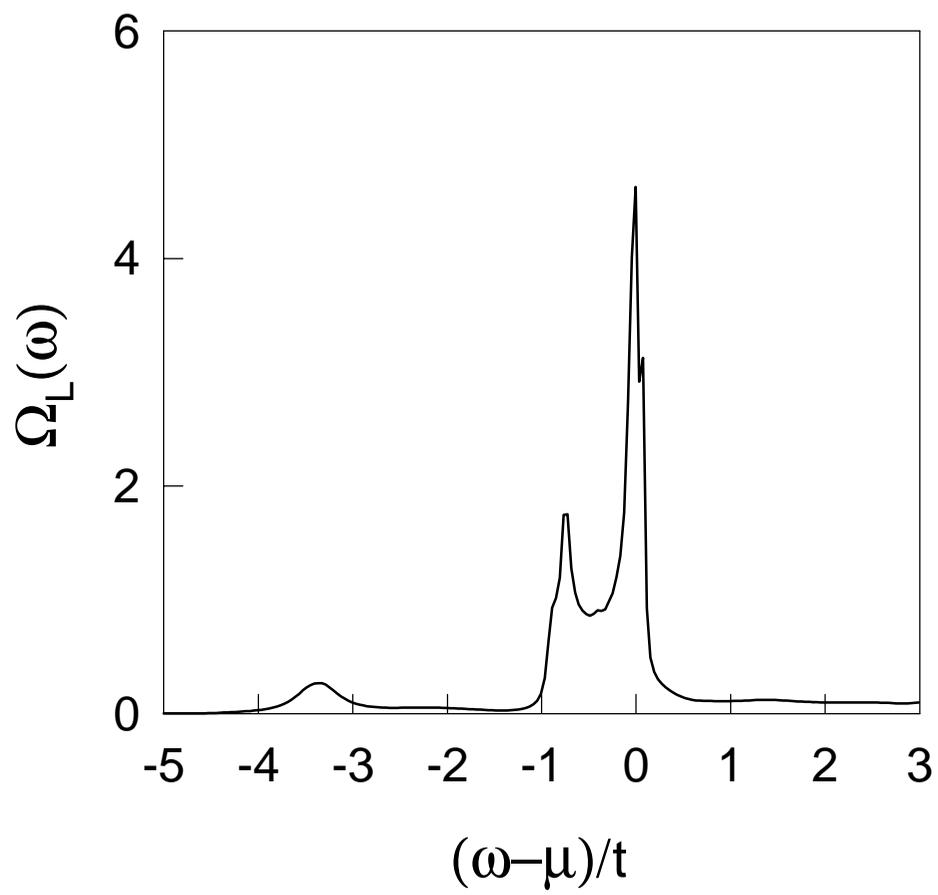
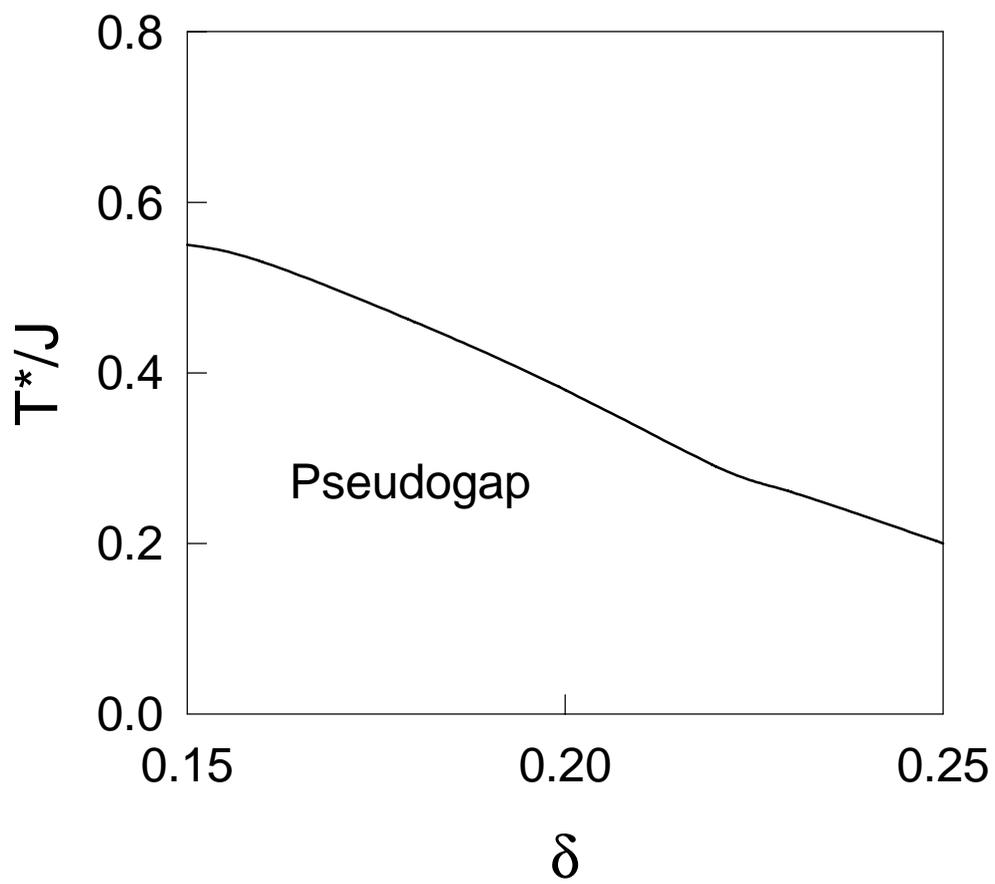
\end{document}